\documentstyle[12pt,graphicx,subfigure]{article}
\def\Journal#1#2#3#4{{#1} {\bf #2}, #3 (#4)}

\def\NPB{{\em Nucl. Phys.} B}
\def\PLB{{\em Phys. Lett.}  B}
\def\PRL{\em Phys. Rev. Lett.}
\def\PRD{{\em Phys. Rev.} D}
\def\ZPC{{\em Z. Phys.} C}

\def\beq{\begin{equation}}
\def\eeq{\end{equation}}
\def\lsim{\ ^<\llap{$_\sim$}\ }
\def\gsim{\ ^>\llap{$_\sim$}\ }
\def\r2{\sqrt 2}
\def\beq{\begin{equation}}
\def\eeq{\end{equation}}
\def\beqn{\begin{eqnarray}}
\def\eeqn{\end{eqnarray}}

\def\sinW2{\sin^2\theta_W}

\def\mz2{M_{z}^2}
\def\c2b{\cos 2\beta}

\def\mz{M_z}

\def\Fq2{F_{2}(q^2)}

\def\sec2w{sec^2\theta_W}
\def\amu{a_\mu}

\def\gmin2{(g-2)_\mu}

\catcode`\@=11 

\def\lsim{\mathrel{\mathpalette\@versim<}}
\def\gsim{\mathrel{\mathpalette\@versim>}}
\def\@versim#1#2{\vcenter{\offinterlineskip
    \ialign{$\m@th#1\hfil##\hfil$\crcr#2\crcr\sim\crcr } }}

\def\PRL{Phys. Rev. Lett.}

\begin{document}
\begin{flushright}
\end{flushright}
\begin{center}
{\Large\bf Interpreting the New Brookhaven  $g_{\mu}-2$ Result\\}
\vglue 0.5cm
{Utpal Chattopadhyay$^{(a)}$ and 
Pran Nath$^{(b)}$
\vglue 0.2cm
{\em 
$^{(a)}$Harish-Chandra Research Institute, Chhatnag Road, Jhusi,\\
Allahabad 211019,India}\\
{\em $^{(b)}$Department of Physics, Northeastern University, Boston,
MA 02115, USA\\} }
\end{center}
\begin{abstract}
The latest $g_{\mu}-2$ measurement by Brookhaven 
confirms the earlier measurement with twice the precision.
However, interpretation of the result requires specific assumptions
regarding the errors in the hadronic light by light (LbL) correction
and in the hadronic vacuum polarization correction. 
Under the assumption that the analysis on LbL
correction of Knecht and Nyffeler and the revised analysis 
of Hayakawa and Kinoshita are
valid the new BNL result implies a deviation 
between experiment and the standard model of $1.6 \sigma -2.6 \sigma$
depending on the estimate of the hadronic vacuum polarization
correction. 
We revisit  the $g_{\mu}-2$ 
constraint for mSUGRA and 
  its implications for the direct detection of 
 sparticles at colliders and for the search for supersymmetric dark matter
 in view of the new evaluation.
\end{abstract}

\section{Introduction}
The BNL $g_{\mu}-2$ Collaboration has announced\cite{bnl2002}
an improved result for $a_{\mu}=(g_{\mu}-2)/2$ with twice the precision of 
their 2001 analysis\cite{bnl2001}.
Here we interpret these
results in the context of  supersymmetry.
It has been recognized for some time that the muon anomalous moment
can act as an important probe of physics beyond the standard model 
especially of supersymmetry. This is so because the new physics 
contribution to the leptonic anomalous magnetic moment 
$a_{\it l}$ scales as $ m_l^2/\Lambda^2$ and 
 is proportional to the square of the lepton mass. 
Thus  $a_{\mu}$  provides a more sensitive probe of  new physics
than $a_e$ even though $a_e$ is more sensitively determined 
while the $a_{\tau}$ determination is
far less sensitive to be competitive. Because of the above any improvement 
in the determination of $a_{\mu}$ has important implications for 
new physics beyond the standard model. Thus
the improved measurement of BNL experiment in 2001\cite{bnl2001} 
which gave $a_{\mu}^{exp}=11659203(15)\times 10^{-10}$ 
resulted in  a large amount of theoretical activity to explore the implications
of the  results.  At the time of the BNL2001 result
 the standard model prediction consisting of
the qed, electro-weak and hadronic corrections was estimated  to be
$a_{\mu}^{SM}= 11659159.7(6.7)\times 10^{-10}$\cite{czar1} 
which gave $a_{\mu}^{exp}- a_{\mu}^{SM}=43(16)\times 10^{-10}$ implying
 a $2.6\sigma$ difference between experiment and theory.
The above difference was based on a  light by light (LbL) hadronic correction
of \cite{hayakawa,bijnens}  $a_{\mu}^{had}(LbL)=-8.5(2.5)\times 10^{-10}$
which was later found to be in 
error\cite{knecht,hkrevised,Bijnens2,Blokland}.  
Thus the analysis of Knecht et.al.\cite{knecht} gives 
 \cite{knecht} $a_{\mu}^{had}(LbL)=8.3(1.2)\times 10^{-10}$
while the revised analysis of Hayakawa and Kinoshita gives\cite{hkrevised} 
$a_{\mu}^{had}(LbL)=8.9(1.5)\times 10^{-10}$. 
A partial analysis of $a_{\mu}^{had}(LbL)$ has also been given in  
Refs.\cite{Bijnens2,Blokland}. Thus the partial analysis of 
 Ref.\cite{Bijnens2} finds 
 $a_{\mu}^{had}(LbL)=8.3(3.2)\times 10^{-10}$  and the work of 
 Ref.\cite{Blokland}  which computed the pion pole part 
 finds $a_{\mu}^{had}(LbL:\pi^0 pole)=5.6\times 10^{-10}$.
Again the sign of these corrections agree with the sign of the
 reevaluation of this quantity in Refs.\cite{knecht,hkrevised}. 
Using the average of the first two\cite{knecht,hkrevised}
which are the more complete calculations and using the hadronic 
vacuum polarization correction of Ref.\cite{davier} one finds
$a_{\mu}^{SM}=11659176.8(6.7)\times 10^{-10}$ and the BNL2001 
result gives 
$a_{\mu}^{exp}-a_{\mu}^{SM}= 26(16)\times 10^{-10}$.
This difference corresponds only to a $1.6\sigma$ deviation between
theory and experiment.

Before proceeding further we wish to discuss a bit further the issue
of errors in the hadronic corrections specifically the LbL correction
and the $\alpha^2$ vacuum polarization correction. Regarding the LbL
hadronic correction, in addition to the analyses mentioned above,
i.e., Refs.\cite{knecht,hkrevised,Bijnens2,Blokland} there is also
the analysis of Ref.\cite{Ramsey} based on chiral 
perturbation theory. This analysis finds 
$a_{\mu}^{had}(LbL) =(1.3^{+5}_{-6}+3.1\tilde C)\times 10^{-10}$ 
where $\tilde C$ is a correction arising from the sub leading 
contributions. These contributions are either of $O(\alpha^3 p^2/\Lambda^2)$,
where p is a mass of order $m_{\mu}$ or $m_{\pi}$ and $\Lambda$ is 
a hadronic scale $\sim 1$ GeV, which are not enhanced by a factor of 
$N_C$ (the number of quark colors) or of  $O(N_C\alpha^3 p^2/\Lambda^2)$
but are not enhanced by large logarithms. The coefficients $\tilde C$
 in this analysis is an  unknown low energy constant (LEC) which 
contains the contributions of nonperturbative physics at short distance.
The authors of Ref.\cite{Ramsey} view the parameter $\tilde C$ as 
basically unconstrained  except through the measurement of the
 anomalous moment itself.
  In our analysis we assume the validity of the analyses of 
  Refs.\cite{knecht,hkrevised}.
 Aside from the LbL contribution, the other source of error  in
 the hadronic correction is the contribution from the $\alpha^2$
 hadronic vacuum polarization 
 correction\cite{davier,narison,yndurain,hadronic}
 and this error is an important component in extracting the
 deviation between experiment and the standard model.

The new  BNL result gives the world average on $a_{\mu}$ as \cite{bnl2002}.  
\beq  
a_{\mu}^{exp}=(11659203)(8)\times 10^{-10}
\eeq
With the LbL correction of Refs.\cite{knecht,hkrevised} and 
assuming the leading hadronic correction to lie in the
range $692(6)\times 10^{-10}$ to 
$702(8)\times 10^{-10}$\cite{davier,narison,yndurain} the
standard model prediction lies in the range 
$a_{\mu}^{SM}=11659177(7)\times 10^{-10}$ to
$a_{\mu}^{SM}=11659186(8)\times 10^{-10}$ 
which leads to 
\beqn
a_{\mu}^{exp}-a_{\mu}^{SM}=(26)(10)\times 10^{-10}\nonumber\\
~~~~~~~~~~~~~~~~~~~~to~~(17)(11)\times 10^{-10}
\eeqn
which corresponds to a difference between experiment and theory
of about $1.6\sigma$ to $2.6\sigma$.

\section{Interpreting the New BNL Result for Supersymmetry}
Next we discuss the implications of these results for 
supersymmetry.
 The  supersymmetric electroweak contribution to $a_\mu$ arises from the 
chargino and neutralino exchange corrections. For the CP conserving
case the chargino ($\chi^{\pm}$) and the sneutrino ($\tilde \nu$)
exchange  contributions are typically the larger contributions
and here one finds\cite{yuan} 
\beq
 a_{\mu}^{\tilde \chi^{\pm}}=\sum_{a=1,2}({{m^2_\mu} \over {48{\pi}^2}}
  {{ {A_R^{(a)}}^2} \over
{m_{\tilde{\chi}_a^\pm}^2}}F_1(\left({{m_{\tilde \nu}} 
\over {m_{\tilde \chi^{\pm}_a}}}\right)^2)+
{{m_\mu} \over{8{\pi}^2}} {{A_R^{(a)}
A_L^{(a)}} \over {m_{\tilde \chi^\pm_a}}} 
F_2(\left({{m_{\tilde \nu}} \over {m_{\tilde \chi^{\pm}_a}}}\right)^2))
\eeq
where 
$A_L(A_R)$ are the left(right) chiral amplitudes and $F_1, F_2$
are form factors as defined in Ref.\cite{yuan}.
The accuracy of the result of Ref.\cite{yuan} 
was further tested by
taking the supersymmetric limit of the results of Ref.\cite{yuan}
 in Ref.\cite{incp} where in addition the effects
of CP violation on $g_{\mu}-2$ were also investigated. 
 The amplitude of Eq.(3) is
dominated by the chiral interference term proportional to $A_LA_R$.
Since $ A_L\sim 1/\cos\beta$ ($\tan\beta =<H_2>/<H_1>$ 
where $<H_2>$ gives mass to the up quark and $<H_1>$ gives mass to the
down quark and the lepton) 
one finds that $\amu^{SUSY}\sim \tan\beta$
for large $\tan\beta$\cite{lopez,chatto}. 
 Further, one finds that $A_L$ 
depends on the sign of $\mu\tilde m_2$,  
where $\tilde m_2$ is $SU(2)$ gaugino mass 
and $\mu$ is the higgs mixing parameter both taken at the electroweak scale,
and thus
the sign of $\amu^{SUSY}$ is determined by the sign of $\mu\tilde m_2$.
The above leads one to the result that over a large  part of the
parameter space unless $\tan\beta \sim 1$ one  finds\cite{lopez,chatto} 
\beq
\amu^{SUSY}>0, \tilde m_2\mu>0; ~~\amu^{SUSY}<0, \tilde m_2\mu<0
\eeq
 where we have used  the sign convention on $\mu$ of Ref.\cite{sugra}.
These results have been confirmed  numerically in a wide class of
models\cite{lopez,chatto,cgr}.
The implications of the BNL result of 2001 has been analyzed
extensively in the 
literature\cite{chatto2,Czarnecki:2001pv,baltz,ccnyuk,icn}
within the framework of mSUGRA\cite{msugra} as well as in a variety of
other models. 

We analyze now the implications of the new result 
within mSUGRA\cite{msugra}. 
We note in passing that for a class of models with large extra dimensions the
correction to g-2 from Kaluza-Klein modes is rather  
small\cite{ny1} once the
constraints arising from Kaluza-Klein corrections to $G_F$ are taken 
account of\cite{ny1}. Thus the extra dimensions do not interfere
in the extraction of implications of the new result for supersymmetry.
In our analysis we will use the $2.6\sigma$ deviation in Eq.(2) 
as the default value for the 
deviation but we will take a $2\sigma$ error corridor in the analysis.
This error  corridor includes the case corresponding to the smaller
 value of the deviation in Eq.(2), i.e., $1.6\sigma$ and  
 taking a $1\sigma$ error corridor in the analysis. 
 Our assumption thus gives us the constraint
      \beq
        6\times 10^{-10} \leq  (a_{\mu}^{exp} - a_{\mu}^{SM})
        \leq 46 \times 10^{-10} 
        \eeq
Attributing the entire difference 
$a_{\mu}^{exp}-a_{\mu}^{SM}$ to supersymmetry and using Eqs.(2) and 
(4) we find that the sign of $\mu$ is positive confirming 
a similar result arrived at in analyses based on the BNL2001 
data\cite{chatto2,Czarnecki:2001pv,baltz}.  
 The sign of $\mu$ is of great importance  for 
 dark matter\cite{baltz,ccnyuk} and for  
 $b-\tau$ unification\cite{bf,ccnyuk}.
As is well known 
$b\rightarrow s+\gamma$ imposes an important constraint on the
parameter space of  supersymmetric
 models\cite{bsgamma,bsgammanew}. 
The constraint is 
very stringent for negative $\mu$ eliminating most of the parameter
space of models. For positive $\mu$ the constraint is not very
strong for small $\tan\beta$ but becomes a strong constraint for
large values of $\tan\beta$. The imposition of this constraint
involves the standard model value and the experimental value
both of which have significant errors. The standard model prediction
of this decay including the next to the leading order 
correction has  been given by several authors\cite{gambino}.
Further, there are several recent experimental determinations
of this decay\cite{cleo}. In our analysis we take a $2\sigma$ 
error corridor  between experiment and the  prediction of 
the standard model and thus allow  the decay branching ratio to 
lie in the range 
$2\times 10^{-4}<  B(b\rightarrow s+\gamma)<4.5\times 10^{-4}$.
Another important aspect of mSUGRA is that it 
leads to the lightest neutralino to be the lightest supersymmetric 
particle (LSP) using the renormalization group analysis
(see, e.g., Ref.\cite{an92}) 
and hence with R parity invariance a candidate for cold dark matter. 
The current astrophysical
data allows the amount of cold dark matter to lie in the range
$0.1<\Omega_{CDM}h^2< 0.3$\cite{Bahcall:1999xn}
 where $\Omega_{CDM}=\rho_{CDM}/\rho_c$
where $\rho_{CDM}$ is the matter density due to cold dark matter and
$\rho_c$ is the critical matter  density needed to close the universe,
and $ h $ is the Hubble parameter in units of 100km/sMpc. In our analysis
we assume the entire cold dark matter as arising from the
lightest neutralino and thus impose the constraint on the neutralino
relic density so that  $0.1<\Omega_{\chi^0}h^2< 0.3$.

The result of the analysis is exhibited in Fig.1 in the $m_0-M_{\frac{1}{2}}$
plane, where $m_0$ is the universal scalar mass and $M_{\frac{1}{2}}$ is
the universal gaugino mass, for $\mu$ positive and for values of 
$\tan\beta$ of 5,10,30 and 50. 
Thus yellow region-I at the top right hand side
is disallowed either due to the absence of EWSB or due to the lighter 
chargino mass lying 
below the current experimental lower limit. Further,  yellow region-II 
 at the bottom is disallowed due to stau becoming the LSP. 
The red area on the boundary between the yellow and the white
region is the zone satisfying the relic density constraint 
$0.1<\Omega_{\chi^0}h^2< 0.3$ (the blue filled circles are the
additional allowed points if the relic density constraint includes  
the lower region $0.02<\Omega_{\chi^0}h^2< 0.1$). 
 As discussed above $b\rightarrow s+\gamma$ is not a strong 
constraint for positive $\mu$ and small $\tan\beta$. As a consequence
 the $b\rightarrow s+\gamma$ constraint does not eliminate any 
 parameter space for the cases $\tan\beta =5$ and $\tan\beta =10$ 
as can be seen from Figs.(1a) and (1b). However, for large values of
$\tan\beta$ the $b\rightarrow s+\gamma$ constraint becomes important
even for positive $\mu$ as can be seen from Figs.(1c) and (1d).
 We turn now to the imposition of the $a_{\mu}^{exp}-a_{\mu}^{SM}$ 
 constraint of Eq.(2).
 In the theoretical computation of the supersymmetric electroweak correction
 we take into account the full one loop supersymmetric contribution
 including the chargino-sneutrino and the neutralino-smuon exchanges.
  The result of imposing $2\sigma$ variation
 around the central value gives the corridor shown in Figs.1 by
 the upper ($ -2\sigma$) and the lower  ($ + 2\sigma$) solid lines.
Specifically the ranges of sparticle
 masses, Higgs boson masses and the $\mu$ parameter 
corresponding to $1\sigma$ and $2\sigma$
 constraints are exhibited in Fig.2. The spectra span a wide range
 and with the lower end of the scale corresponding to a low 
 values of fine tuning and the high end corresponding to high 
 values of fine tuning (see, e.g., Ref.\cite{ccn}). 
 The spectrum of Fig.2 leads us to the conclusion
 that sparticles satisfying  $1\sigma$ constraint are
 well within reach of the LHC\cite{cms} and some of the
 sparticles could also show up at RUNII of the Tevatron.
 The ranges allowed by the $2\sigma$ constraint constraint have
 upper limits which in some cases reach as high as 3 TeV and may 
 exceed the reach of the LHC. However,  Fig. 2 does not contain
 the constraints of $b\rightarrow s+\gamma$ or of relic density.
 With these additional constraints the upper limits would be substantially
 smaller. 
 
 We discuss now the direct detection of dark matter under the 
 BNL2002 $g_{\mu}-2$ constraint.
 The analysis of Fig.1 shows that there are substantial regions
 of the parameter space where all the constraints, i.e., the g-2 constraint
 the $b\rightarrow s+\gamma$ constraint and the relic density constraints
 are simultaneously satisfied. We discuss now the direct detection of 
 dark matter under these constraints.
  There has been considerable 
 work on the theoretical predictions of the neutralino-proton
 cross section $\sigma_{\chi p}$ which enters in the direct 
 detection of dark matter and some recent literature
 can be found in  Ref.\cite{cross}. Here we give the
 analysis of these cross sections in mSUGRA for positive $\mu$ 
 under the new $g_{\mu}-2$ constraint. Results are exhibited in
 Figure 3 for values of $\tan\beta$ of 5,10,30 and 50. 
 We find that the neutralino-proton cross sections consistent
 with all the constraints lie in the range  
 $10^{-46} cm^2\leq\sigma_{\chi p}\leq2\times 10^{-43} cm^2$. 
 A significant part of this range will
 be accessible  at future dark matter detectors GENIUS\cite{genius} 
 and ZEPLIN\cite{zeplin}.

\section{Conclusion}     
 The new Brookhaven measurement of $g_{\mu}-2$ has confirmed the
 earlier measurement with  twice the precision. However, 
 interpretation of what this implies for physics beyond the standard
 model is very sensitive to the errors in the standard model 
 prediction. The main surprise at the end of last year was the
 switch in sign of the LbL contribution. There is now a general agreement
 that the sign of this contribution is positive and also an
  agreement between two independent evaluations\cite{knecht,hkrevised} 
  on its size. An important exception to this is the result of
 chiral perturbation theory, which although a more fundamental approach is 
 beset by the appearance of unknown low energy constants which require
 a nonperturbative approach such as lattice gauge calculation 
 for its evaluation. In the present analysis we have assumed the
 validity of the analysis  on LbL  of  Knecht and Nyffeler and the 
 revised analysis of  Hayakawa and Kinoshita.
  Taking account of all the errors the difference between experiment
 and the standard model is now predicted to lie between $1.6\sigma$
 and $2.6\sigma$. We have revisited the implications of the 
 $g_{\mu}-2$ constraint for supersymmetry in view of the new 
 result. Using the $2.6\sigma$ difference and a $2\sigma$ error 
 corridor our analysis points to significant regions in the mSUGRA
  parameter space
  consistent with the $b\rightarrow s+\gamma$, relic density and
  $g_{\mu}-2$ constraints. Within $1\sigma$ error corridor
  all of the sparticles are accessible at the Large Hadron
  collider.  Further, for low values
  of $\tan\beta$ 
  sparticles may also be  accessible at Fermilab Tevatron.
   We also carried out an analysis of 
  the  neutralino-proton cross section $\sigma_{\chi^0-p}$ and
  find that a significant part of the parameter space would be
  explored by the future dark matter detectors such as
  CDMS (Soudan), GENIUS and ZEPLIN. Finally, the $g_{\mu}-2$
  experiment could lead to an even more stringent constraint if
  there was a reduction in the error associated with the
  standard model prediction.
  This could come about by eliminating the ambiguity on the LbL
  correction through a lattice gauge calculation analysis, 
  and through improved low energy data on 
  $e^+e^-\rightarrow hadrons$ in the analysis of hadronic 
  vacuum polarization correction.

~\\
We thank William Marciano for an interesting discussion.  
This work is  supported in part by NSF grant PHY-9901057.\\

\newpage
\noindent
Figure Captions\\
Figure 1: Analysis in mSUGRA  exhibiting the allowed and the 
disallowed regions in the $m_0-M_{\frac{1}{2}}$ plane 
due to the  radiative breaking of the electro-weak symmetry,
the $b\rightarrow s+\gamma$ constraint and the 
neutralino relic density constaint as discussed in the text for
values of $\tan\beta$ of 5,10,30 and 50. 
The area to the right of the dotted
 blue line in each case is the allowed region by the 
$b\rightarrow s+\gamma$ constraint and the (red) dark area satisfy
the relic density constraint $0.1<\Omega_{\chi^0}h^2< 0.3$
(while  the (blue) dark filled circles satisfy 
$0.02<\Omega_{\chi^0}h^2< 0.1$).
The region I is discarded because of the absence of 
the electroweak radiative symmetry breaking or via the 
lower limit of chargino mass constraint.  The region II is eliminated 
because of stau becoming the LSP.
 The solid black lines exhibit the lower limit
($a_{\mu}(2\sigma)$) and the upper ($a_{\mu}(-2\sigma)$) 
limit corresponding to the $\pm 2\sigma$ variation around
the central value given by the first line of Eq.(2).
The LEP 'higgs signal' is also indicated in the figures.\\

\noindent
Figure 2: 
Mass ranges of sparticles and Higgs bosons as well as the $\mu$ 
parameter in mSUGRA for $\mu>0$  
when the $1\sigma$ and
$2\sigma$ $g_{\mu}-2$ constraints are imposed and $\tan\beta$
ranges from 2-60 and $A_0=0$. The 
 $b\rightarrow s+\gamma$ constraint and the
relic density constraint are
not imposed. For each (s)particle the vertical line to the left is the range
in mass allowed by a $2\sigma$ $g_{\mu}-2$ constraint while the 
vertical line to the right is the mass range allowed by a 
   $1\sigma$ $g_{\mu}-2$ constraint.\\

\noindent
Figure 3: 
A plot of the scalar proton-LSP cross-section $\sigma_{\chi p}$ vs LSP mass
$m_{\chi}$ which enters in the direct detection rates.
The blue dots are the points allowed in mSUGRA and the red filled circles 
satisfy the constraint $0.1< \Omega h^2< 0.3$. 
Black squares which form the black region 
satisfy the $2\sigma$ constraints 
on  $b\rightarrow s+\gamma$ and  $g_{\mu}-2$.
The enclosed region on the top left hand side is the region where the
DAMA collaboration\cite{dama} claims a signal. The dot-dashed line on top is 
the upper limit from CDMS\cite{cdms} while the solid line is the
 latest upper limit from EDELWEISS experiment\cite{edelweiss}.
 The lower dashed line is the sensitivity that will be reached by the
 CDMS experiment at Soudan mine while the dotted line is the sensitivity
 that will be achieved by the GENIUS detector\cite{genius}.

\newpage
\begin{figure}           
\vspace*{-1.0in}                                 
\subfigure[]{                       
\label{amutan5} 
\hspace*{-0.6in}                     
\begin{minipage}[b]{0.5\textwidth}                       
\centering
\includegraphics[width=\textwidth,height=\textwidth]{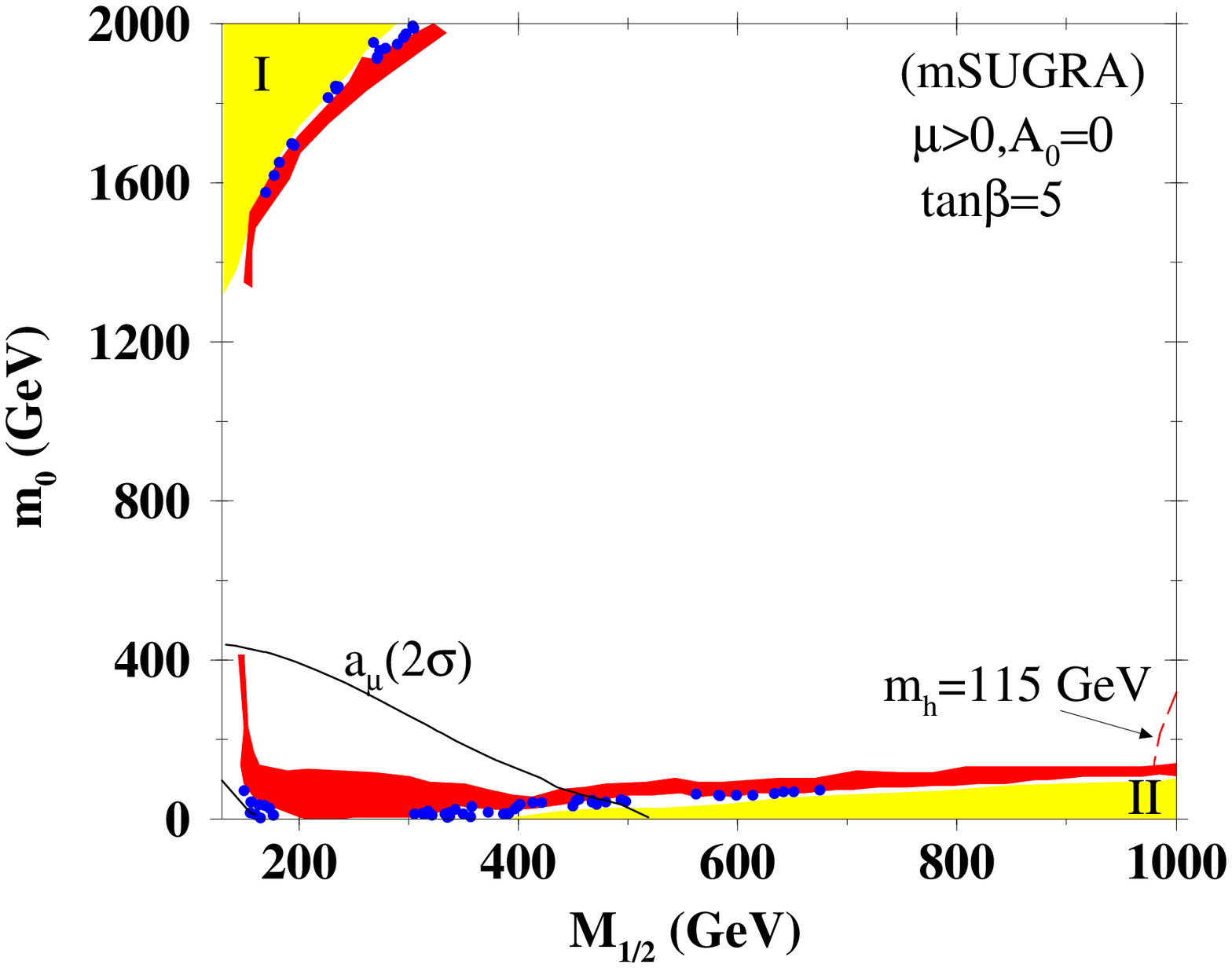}    
\end{minipage}}                       
\hspace*{0.3in}
\subfigure[]{      
\label{amutan10}                  
\begin{minipage}[b]{0.5\textwidth}                       
\centering                      
\includegraphics[width=\textwidth,height=\textwidth]{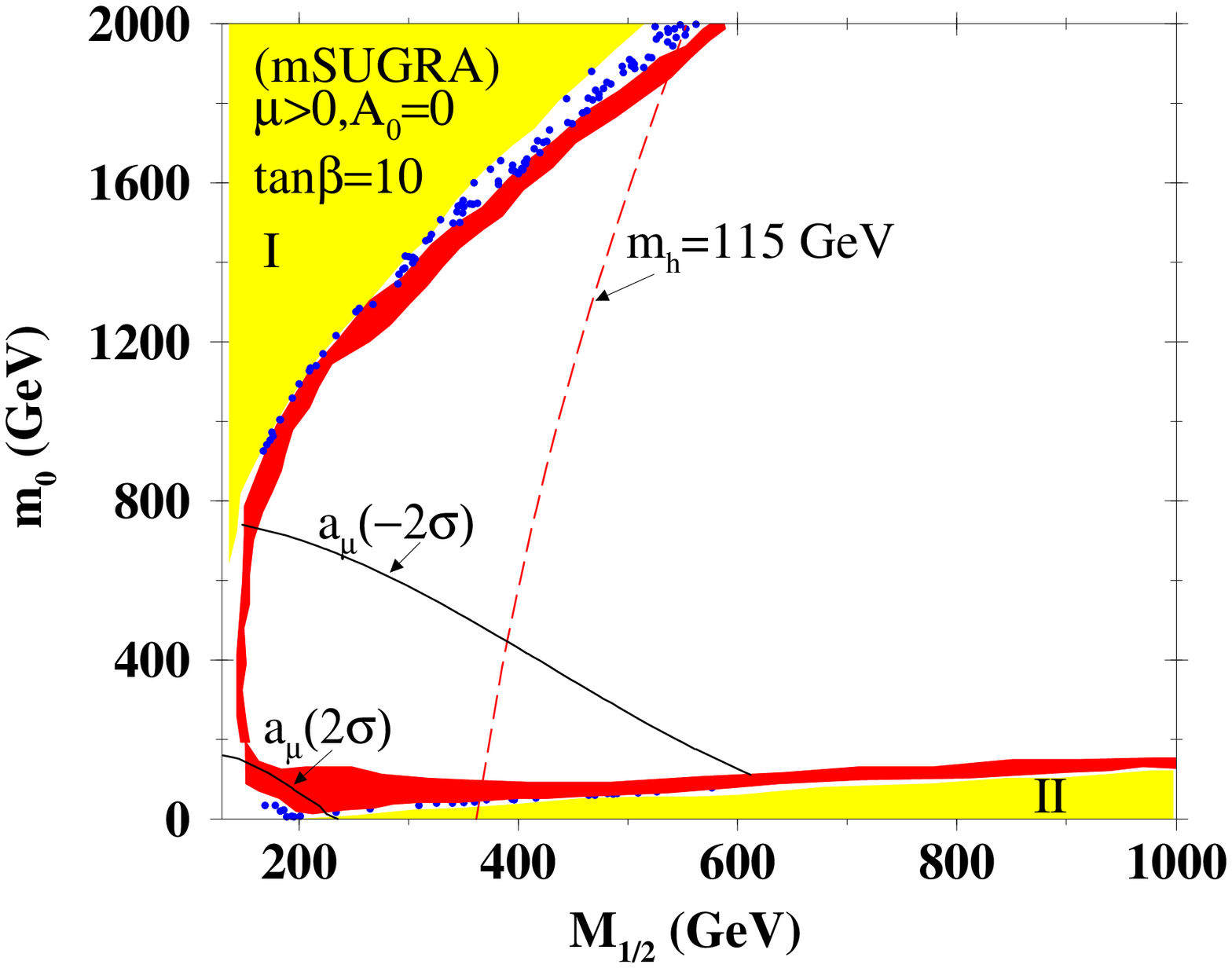} 
\end{minipage}}                       
\hspace*{-0.6in}                     
\subfigure[]{                       
\label{amutan30}                  
\begin{minipage}[b]{0.5\textwidth}                       
\centering
\includegraphics[width=\textwidth,height=\textwidth]{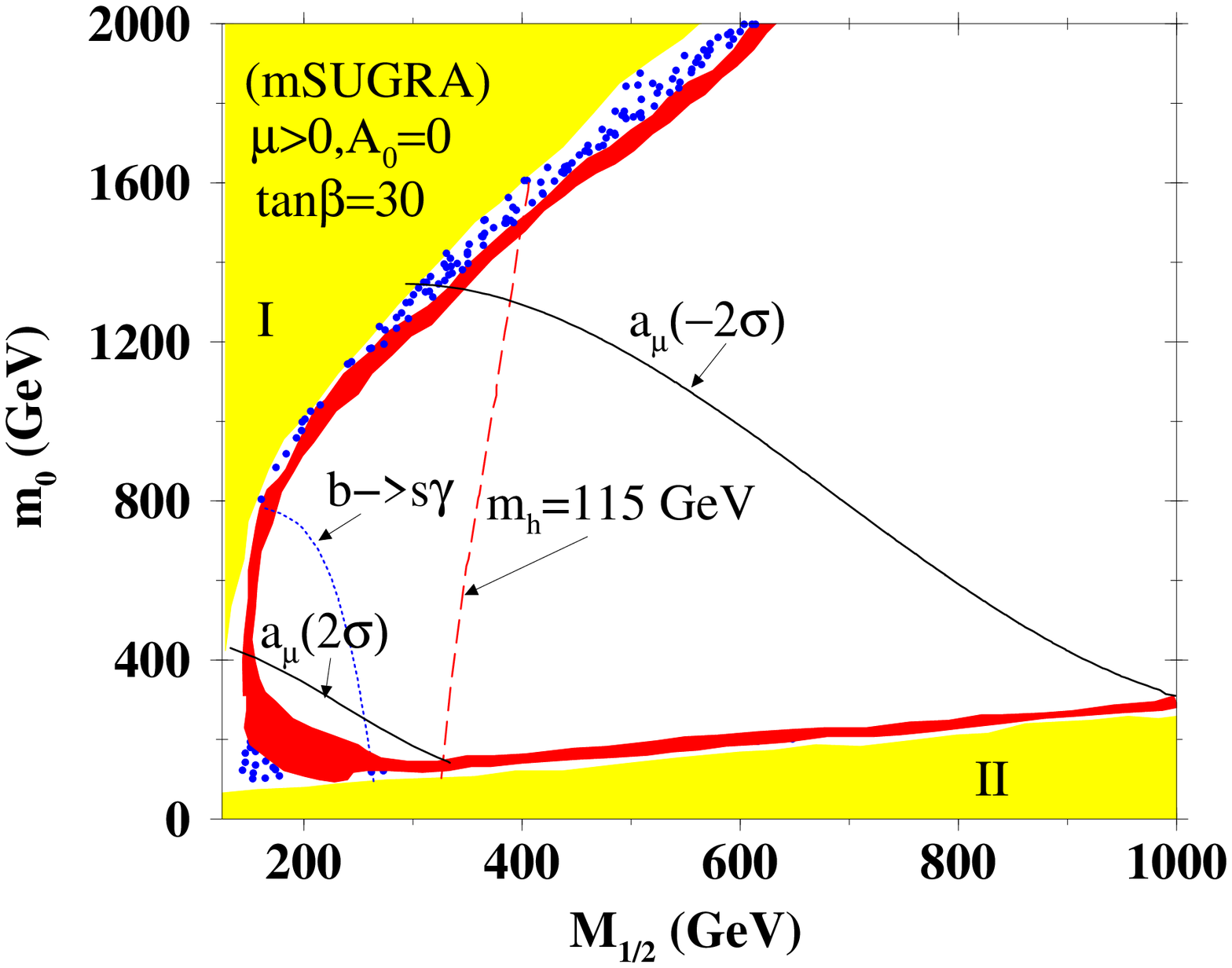}
\end{minipage}}
\hspace*{0.3in}                       
\subfigure[]{                       
\label{amutan45}
\begin{minipage}[b]{0.5\textwidth}                       
\centering                      
\includegraphics[width=\textwidth,height=\textwidth]{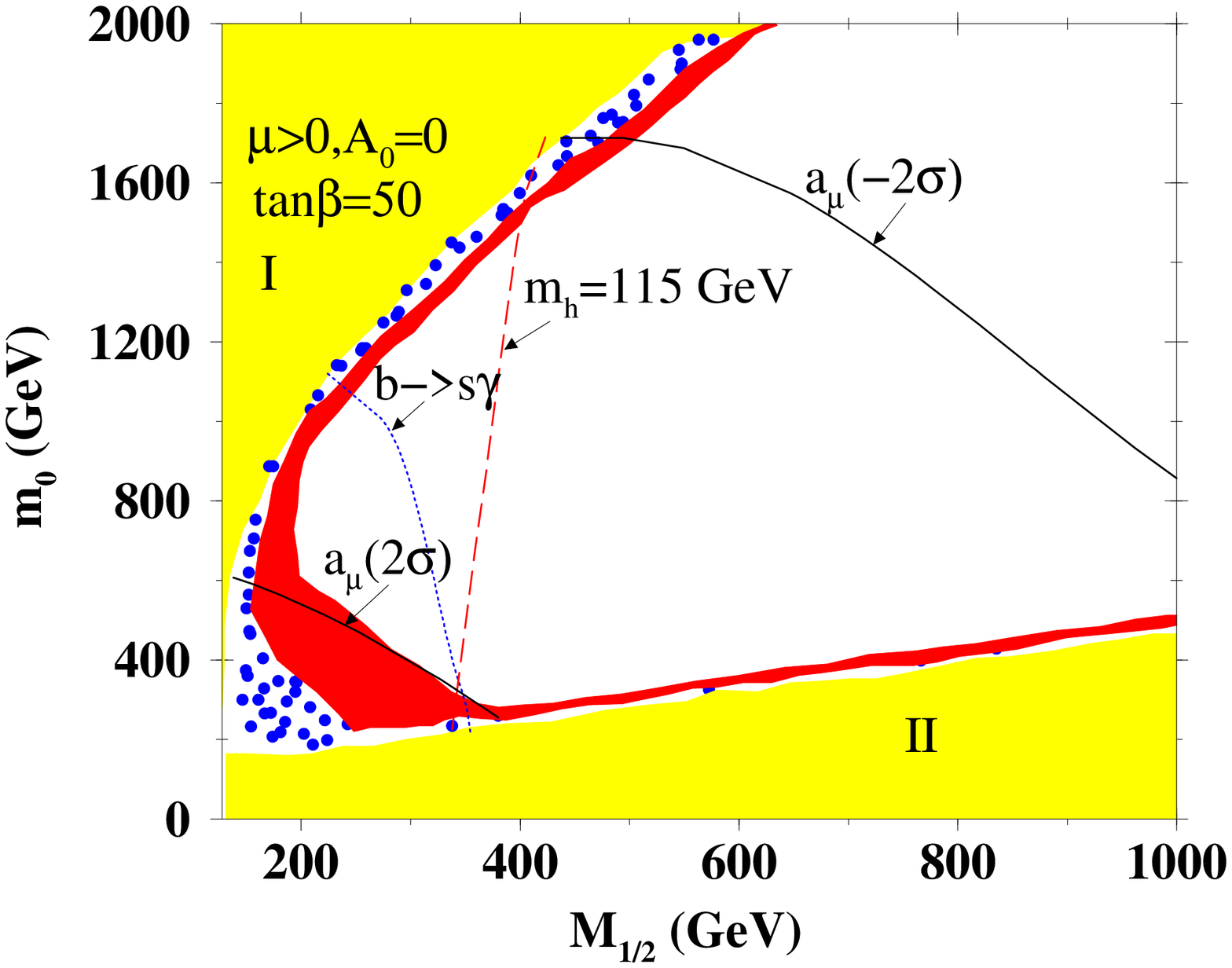}
\end{minipage}}
\caption{}
\label{amutan}  
\end{figure}

\newpage
\begin{figure}
\centering 
\includegraphics[width=\textwidth,height=\textwidth]{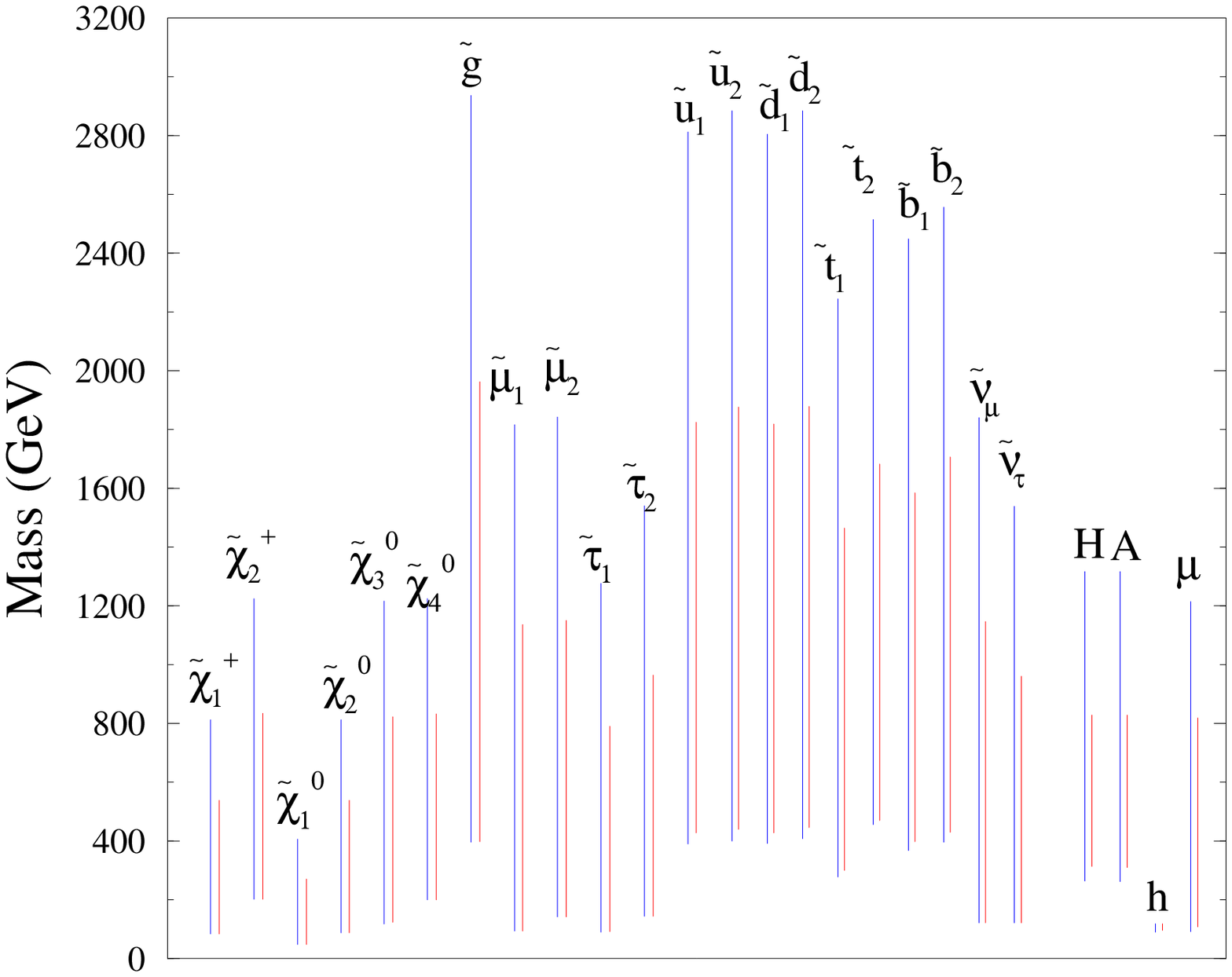}
\caption{}
\label{spectra} 
\end{figure}

\newpage
\begin{figure}           
\vspace*{-1.0in}                                 
\subfigure[]{                       
\label{xamutan5} 
\hspace*{-0.6in}                     
\begin{minipage}[b]{0.5\textwidth}                       
\centering
\includegraphics[width=\textwidth,height=\textwidth]{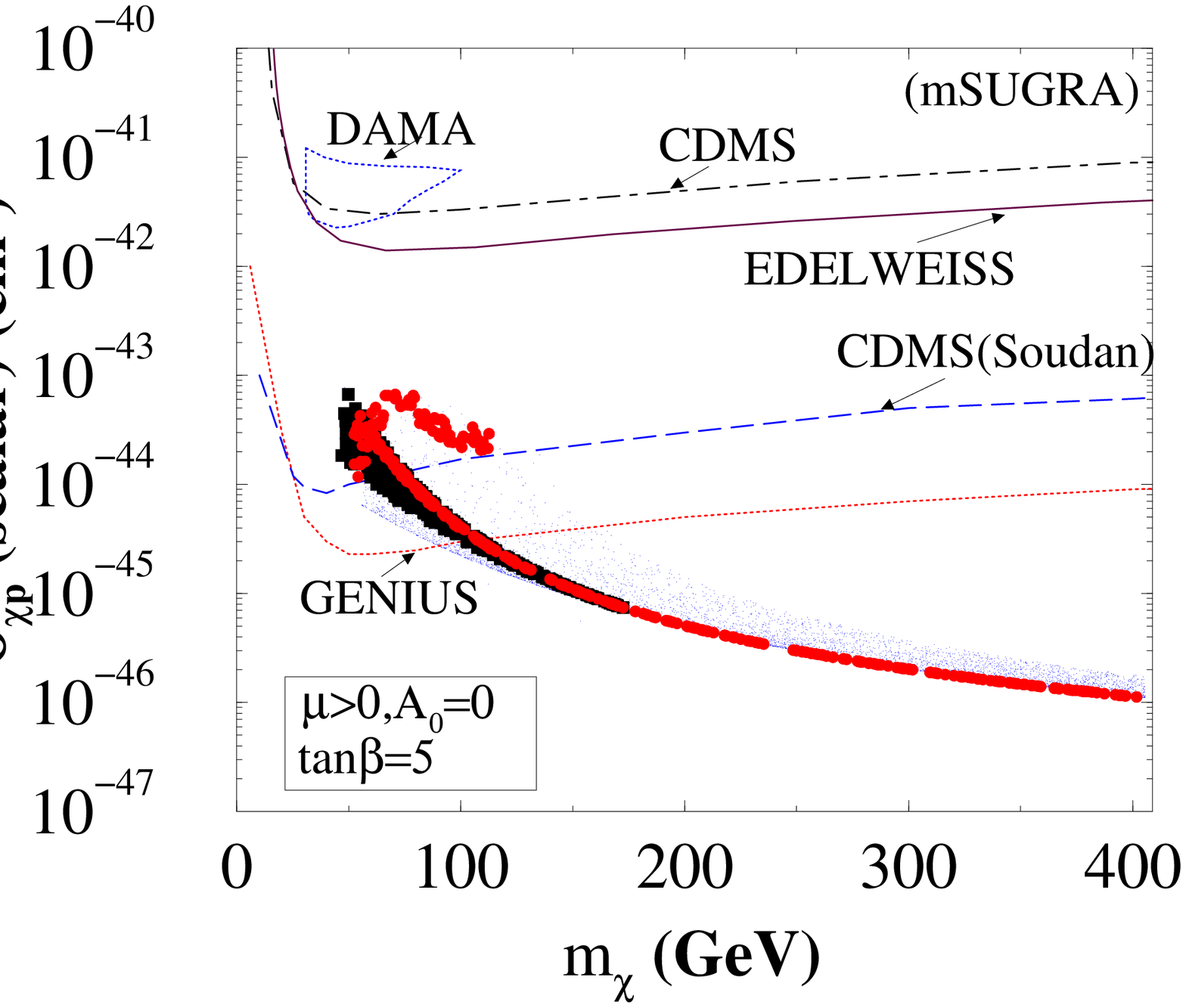}    
\end{minipage}}                       
\hspace*{0.3in}
\subfigure[]{      
\label{xamutan10}                  
\begin{minipage}[b]{0.5\textwidth}                       
\centering                      
\includegraphics[width=\textwidth,height=\textwidth]{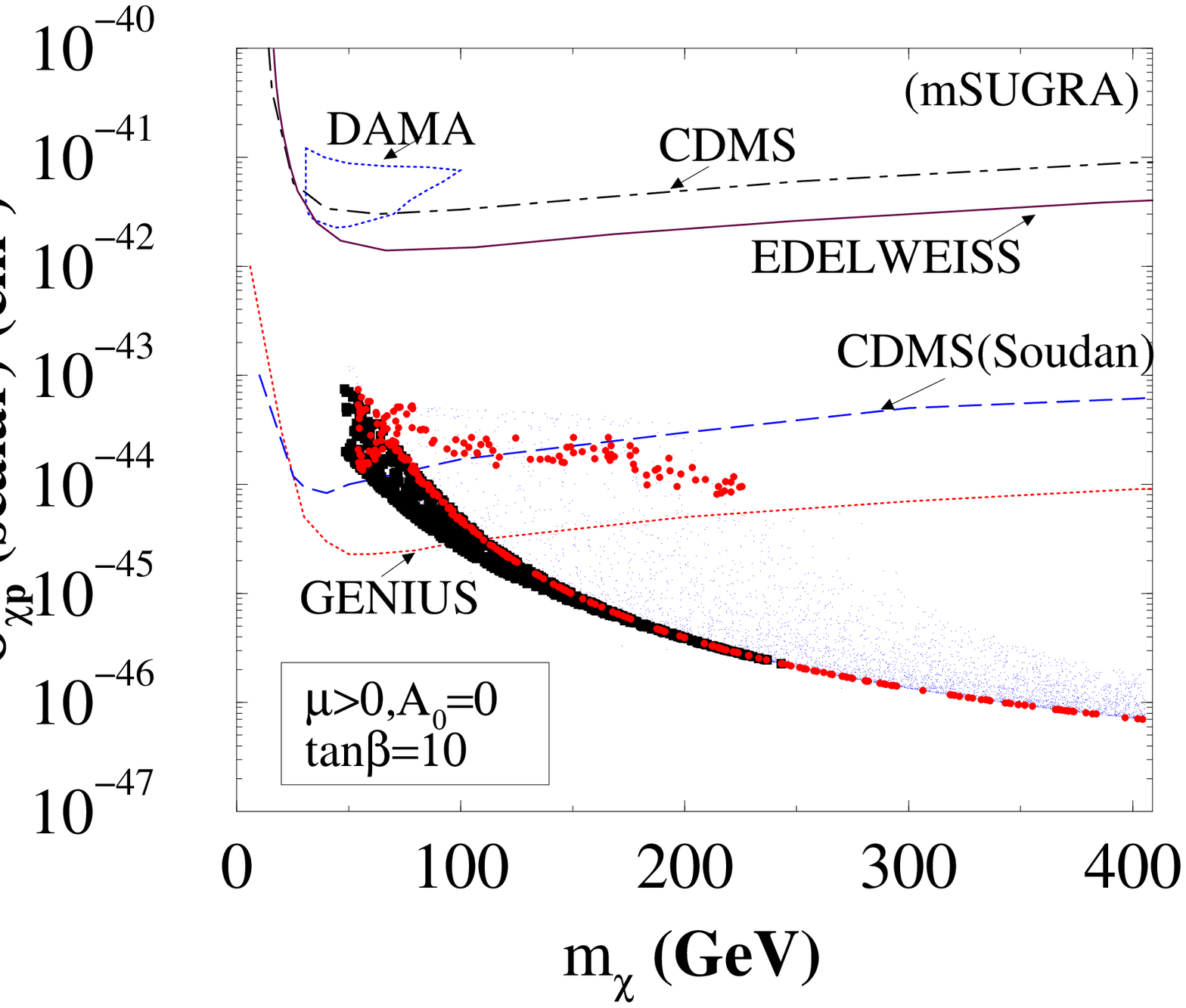} 
\end{minipage}}                       
\hspace*{-0.6in}                     
\subfigure[]{                       
\label{xamutan30}                  
\begin{minipage}[b]{0.5\textwidth}                       
\centering
\includegraphics[width=\textwidth,height=\textwidth]{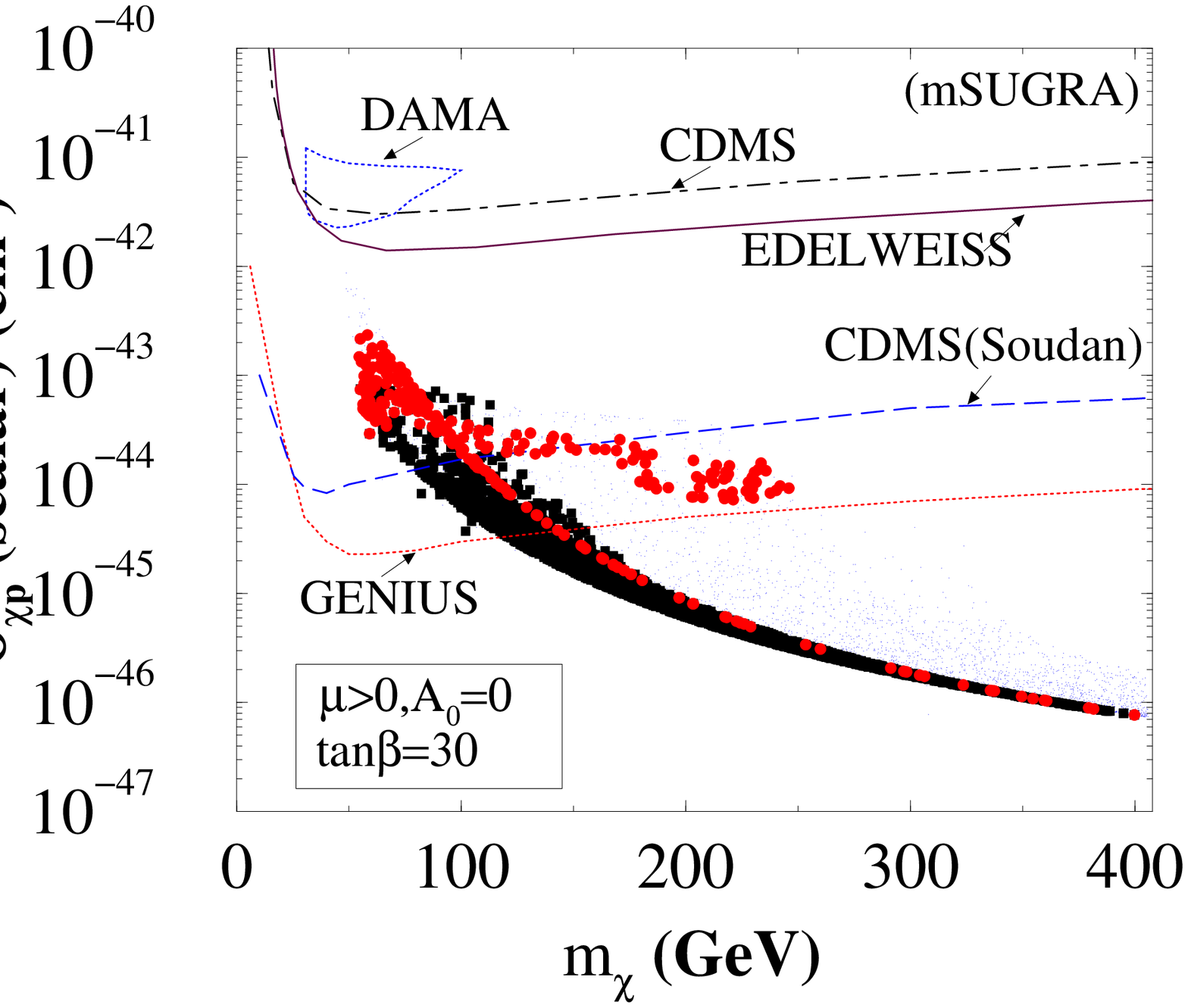}
\end{minipage}}
\hspace*{0.3in}                       
\subfigure[]{                       
\label{xamutan45}
\begin{minipage}[b]{0.5\textwidth}                       
\centering                      
\includegraphics[width=\textwidth,height=\textwidth]{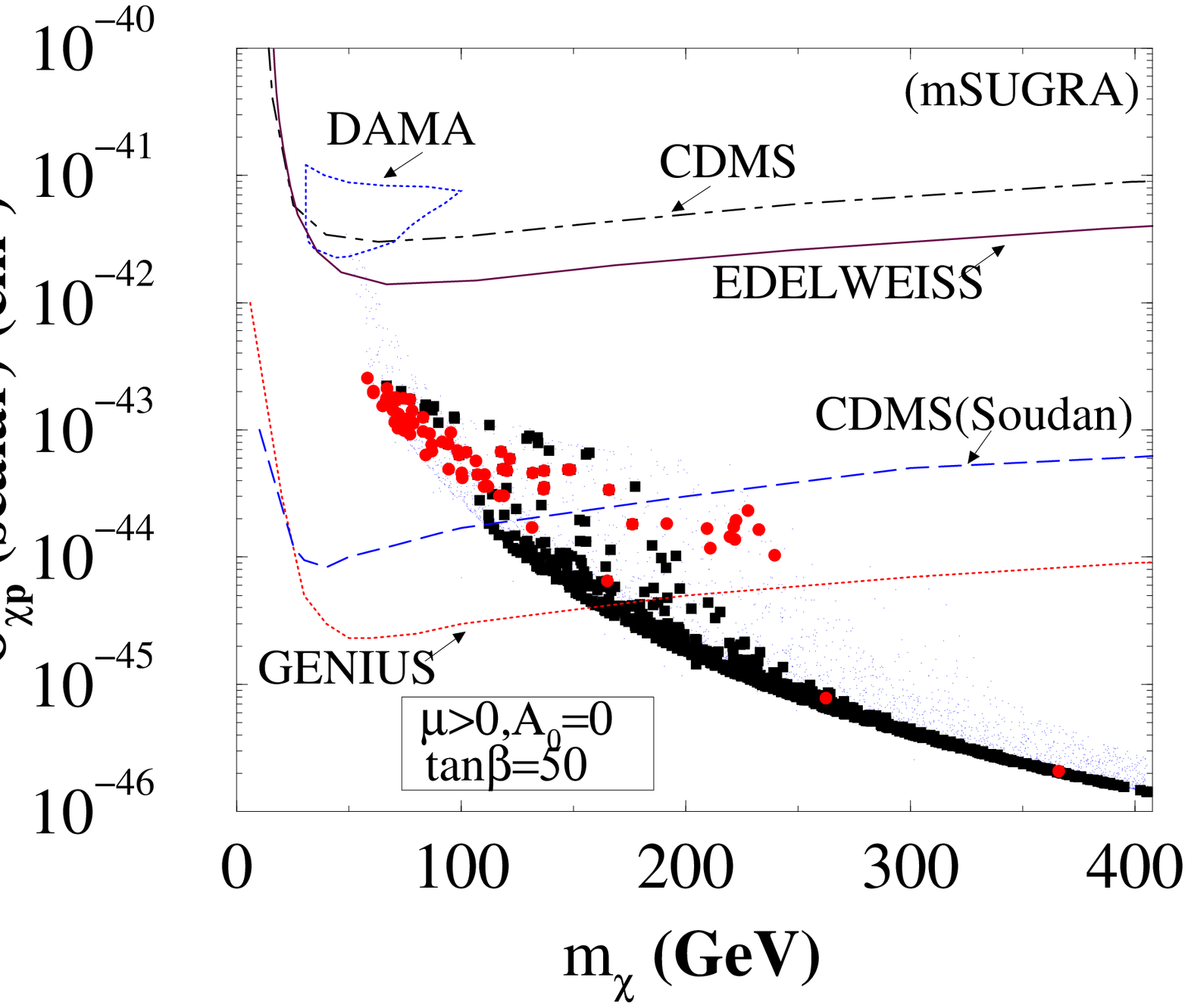}
\end{minipage}}
\caption{}
\label{xamutan} 
\end{figure}


\begin{thebibliography}{999}

\bibitem{bnl2002}
G.W. Bennet {\it et al.}  [Muon g-2 Collaboration],
hep-ex/0208001



 \bibitem{bnl2001}
H.~N.~Brown {\it et al.}  [Muon g-2 Collaboration],
Phys.\ Rev.\ Lett.\  {\bf 86}, 2227 (2001)
[arXiv:hep-ex/0102017].



\bibitem{czar1}
A.~Czarnecki and W.~J.~Marciano,
Nucl.\ Phys.\ Proc.\ Suppl.\  {\bf 76}, 245 (1999)
[arXiv:hep-ph/9810512].



\bibitem{hayakawa}
M.~Hayakawa, T.~Kinoshita and A.~I.~Sanda,
Phys.\ Rev.\ Lett.\  {\bf 75}, 790 (1995)
[arXiv:hep-ph/9503463].
;
M.~Hayakawa, T.~Kinoshita and A.~I.~Sanda,
Phys.\ Rev.\ D {\bf 54}, 3137 (1996)
[arXiv:hep-ph/9601310].
;
M.~Hayakawa and T.~Kinoshita,
Phys.\ Rev.\ D {\bf 57}, 465 (1998)
[arXiv:hep-ph/9708227].




\bibitem{bijnens}
J.~Bijnens, E.~Pallante and J.~Prades,
Phys.\ Rev.\ Lett.\  {\bf 75}, 1447 (1995)
[Erratum-ibid.\  {\bf 75}, 3781 (1995)]
[arXiv:hep-ph/9505251].
;
J.~Bijnens, E.~Pallante and J.~Prades,
Nucl.\ Phys.\ B {\bf 474}, 379 (1996)
[arXiv:hep-ph/9511388].



\bibitem{knecht}
M.~Knecht and A.~Nyffeler,
Phys.\ Rev.\ D {\bf 65}, 073034 (2002)
[arXiv:hep-ph/0111058].
;
M.~Knecht, A.~Nyffeler, M.~Perrottet and E.~De Rafael,
Phys.\ Rev.\ Lett.\  {\bf 88}, 071802 (2002)
[arXiv:hep-ph/0111059].



\bibitem{hkrevised}
M.~Hayakawa and T.~Kinoshita,
arXiv:hep-ph/0112102.


\bibitem{Bijnens2}
J.~Bijnens, E.~Pallante and J.~Prades,
Nucl.\ Phys.\ B {\bf 626}, 410 (2002)
[arXiv:hep-ph/0112255].


\bibitem{Blokland}
I.~Blokland, A.~Czarnecki and K.~Melnikov,
Phys.\ Rev.\ Lett.\  {\bf 88}, 071803 (2002)
[arXiv:hep-ph/0112117].


\bibitem{davier}
M.~Davier and A.~Hocker,
Phys.\ Lett.\ B {\bf 435}, 427 (1998)
[arXiv:hep-ph/9805470].



\bibitem{Ramsey}
M.~Ramsey-Musolf and M.~B.~Wise,
Phys.\ Rev.\ Lett.\  {\bf 89}, 041601 (2002)
[arXiv:hep-ph/0201297].


\bibitem{narison}
S.~Narison,
Phys.\ Lett.\ B {\bf 513}, 53 (2001)
[Erratum-ibid.\ B {\bf 526}, 414 (2002)]
[arXiv:hep-ph/0103199].

\bibitem{yndurain}
J.~F.~De Troconiz and F.~J.~Yndurain,
Phys.\ Rev.\ D {\bf 65}, 093001 (2002)
[arXiv:hep-ph/0106025].



\bibitem{hadronic}
For other evaluations of the error in $\alpha^2$ vacuum polarization
hadronic corrections see,
K.~Melnikov,
Int.\ J.\ Mod.\ Phys.\ A {\bf 16}, 4591 (2001)
[arXiv:hep-ph/0105267].
;
G.~Cvetic, T.~Lee and I.~Schmidt,
Phys.\ Lett.\ B {\bf 520}, 222 (2001)
[arXiv:hep-ph/0107069].



\bibitem{yuan}
  T. C. Yuan, R. Arnowitt, A. H. Chamseddine and P. Nath, 
 \Journal {\ZPC}{26}{407}{1984};
 D.A. Kosower, L.M. Krauss, N. Sakai, \Journal{\PLB}{133}{305}{1983};

\bibitem{incp}
T.~Ibrahim and P.~Nath,
Phys.\ Rev.\ D {\bf 61}, 095008 (2000)
[arXiv:hep-ph/9907555].
; ibid, 
Phys.\ Rev.\ D {\bf 62}, 015004 (2000)
[arXiv:hep-ph/9908443].



\bibitem{lopez}
J.L. Lopez, D.V. Nanopoulos, X. Wang, \Journal{\PRD}
{49}{366}{1994}.

\bibitem{chatto}
U. Chattopadhyay and P. Nath, \Journal{\PRD}{53}{1648}{1996};
T. Moroi, \Journal{\PRD}{53}{6565}{1996}

\bibitem{sugra}
S.~Abel {\it et al.}  [SUGRA Working Group Collaboration],
arXiv:hep-ph/0003154.

\bibitem{cgr}
T.~Moroi,
Phys.\ Rev.\ D {\bf 53}, 6565 (1996)
[Erratum-ibid.\ D {\bf 56}, 4424 (1997)]
[arXiv:hep-ph/9512396].
;
U.~Chattopadhyay, D.~K.~Ghosh and S.~Roy,
Phys.\ Rev.\ D {\bf 62}, 115001 (2000)
[arXiv:hep-ph/0006049].

\bibitem{chatto2}
U.~Chattopadhyay and P.~Nath,
Phys.\ Rev.\ Lett.\  {\bf 86}, 5854 (2001).

\bibitem{Czarnecki:2001pv}
A.~Czarnecki and W.~J.~Marciano,
Phys.\ Rev.\ D {\bf 64}, 013014 (2001)
[arXiv:hep-ph/0102122]
;
L.~L.~Everett, G.~L.~Kane, S.~Rigolin and L.~Wang, 
Phys.\ Rev.\ Lett.\  {\bf 86}, 3484 (2001);
J.~L.~Feng and K.~T.~Matchev, Phys.\ Rev.\ Lett.\  {\bf 86}, 3480 (2001);
 S. Komine, T. Moroi, and M. Yamaguchi, Phys.\ Lett.\ B {\bf 506}, 93 (2001);
Phys.\ Lett.\ B {\bf 507}, 224 (2001);
S. P. Martin, J. D. Wells, Phys.\ Rev.\ D {\bf 64}, 035003 (2001);
H. Baer, C. Balazs, J. Ferrandis, X. Tata, 
Phys.Rev.{\bf D64}: 035004, (2001); 
M. Byrne, C. Kolda,  J.E. Lennon, arXiv:hep-ph/0108122. 
For a more complete set
of references see, U.~Chattopadhyay and P.~Nath,
arXiv:hep-ph/0108250.



\bibitem{baltz}
E.~A.~Baltz and P.~Gondolo, Phys.\ Rev.\ Lett.\  {\bf 86}, 5004 (2001);
J. Ellis, D.V. Nanopoulos, K. A. Olive, Phys.\ Lett.\ B {\bf 508}, 65 (2001);
R. Arnowitt, B. Dutta, B. Hu, Y. Santoso, 
Phys.\ Lett.\ B {\bf 505}, 177 (2001).


\bibitem{ccnyuk}
U.~Chattopadhyay, A.~Corsetti and P.~Nath,
arXiv:hep-ph/0201001; arXiv:hep-ph/0202275
(to appear in Phys.Rev.D). 

\bibitem{icn}
T.~Ibrahim, U.~Chattopadhyay and P.~Nath,
Phys.\ Rev.\ D {\bf 64}, 016010 (2001)
[arXiv:hep-ph/0102324].



\bibitem{msugra}
A.H. Chamseddine, R. Arnowitt and P. Nath, \Journal{\PRL}{49}
{970}{1982}; R. Barbieri, S. Ferrara and C.A. Savoy, \Journal{\PLB}
{119}{343}{1982}; L. Hall, J. Lykken, and S. Weinberg,
\Journal{\PRD}{27}{2359}{1983}: P. Nath, R. Arnowitt and A.H. Chamseddine,
\Journal{\NPB}{227}{121}{1983}. For reviews, see P. Nath, R. Arnowitt
and A.H. Chamseddine, "Applied N=1 Supergravity", world scientific,
1984; H.P. Nilles, Phys. Rep. {\bf 110}, 1(1984).



\bibitem{ny1}
P.~Nath and M.~Yamaguchi,
Phys.\ Rev.\ D {\bf 60}, 116004 (1999);
Phys.\ Rev.\ D {\bf 60}, 116006 (1999).
See also K.~Agashe, N.~G.~Deshpande and G.~H.~Wu,
Phys.\ Lett.\ B {\bf 489}, 367 (2000).
M.~L.~Graesser,
Phys.\ Rev.\ D {\bf 61}, 074019 (2000)
[arXiv:hep-ph/9902310].




\bibitem{bf}
H. Baer and J. Ferrandis, Phys. Rev. Lett.{\bf 87}, 211803 (2001);
T.~Blazek, R.~Dermisek and S.~Raby,
Phys.\ Rev.\ Lett.\  {\bf 88}, 111804 (2002);  
ibid, Phys.\ Rev.\ D {\bf 65}, 115004 (2002);
S.~Komine and M.~Yamaguchi,
Phys.\ Rev.\ D {\bf 65}, 075013 (2002)
[arXiv:hep-ph/0110032].
;
U.~Chattopadhyay and P.~Nath,
Phys.\ Rev.\ D {\bf 65}, 075009 (2002).


\bibitem{bsgamma} 
P. Nath and R. Arnowitt, \Journal{\PLB}{336}{395}{1994};
\Journal{\PRL}{74}{4592}{1995};
F.~Borzumati, M.~Drees and M.~Nojiri, \Journal{\PRD}{51}{341}{1995};
H. Baer, M. Brhlik, D. Castano and  X. Tata, \Journal{\PRD}{58}
{015007}{1998}.

\bibitem{bsgammanew}
M. Carena, D. Garcia, U. Nierste, C.E.M. Wagner, Phys. Lett. 
{\bf B499} 141 (2001); 
G. Degrassi, P. Gambino, G.F. Giudice, JHEP 0012, 009 (2000).


\bibitem{gambino}
P. Gambino and M. Misiak, Nucl. Phys. {\bf B611}, 338 (2001); 
 P. Gambino and U. Haisch,  JHEP 0110, 020 (2001);
 A.L. Kagan and M. Neubert, Eur. Phys. J. C7, 5(1999).
A.L. Kagan and M. Neubert, Eur. Phys. J. {\bf C27}, 5(1999).  

\bibitem{cleo}
S. Chen et.al. (CLEO Collaboration), Phys. Rev. Lett. {\bf 87}, 251807 
(2001);
H. Tajima, talk at the 20th International Symposium on 
Lepton-Photon Interactions", Rome, July 2001;
R. Barate et.al., Phys. Lett. {\bf B429}, 169(1998).


\bibitem{an92}
Arnowitt and P.~Nath,
Phys.\ Rev.\ Lett.\  {\bf 69}, 725 (1992);
P.~Nath and R.~Arnowitt,
Phys.\ Lett.\ B {\bf 287}, 89 (1992).


\bibitem{Bahcall:1999xn}
N.~A.~Bahcall, J.~P.~Ostriker, S.~Perlmutter and P.~J.~Steinhardt,
Science {\bf 284}, 1481 (1999)
[arXiv:astro-ph/9906463].


\bibitem{ccn}
K.L. Chan, U. Chattopadhyay and P. Nath, \Journal {\PRD}{58}{096004}{1998}.

\bibitem{cms}
CMS Collaboration, Technical Proposal: CERN/LHCC 94-38(1994);
ATLAS Collaboration, Technical Proposal, CERN/LHCC 94-43(1944);
H.~Baer, C.~h.~Chen, F.~Paige and X.~Tata,
Phys.\ Rev.\ D {\bf 52}, 2746 (1995)
[arXiv:hep-ph/9503271].


\bibitem{cross}
J.~R.~Ellis, A.~Ferstl and K.~A.~Olive,
arXiv:hep-ph/0106148.
;arXiv:hep-ph/0106148;
A.~Corsetti and P.~Nath,
Phys.\ Rev.\ D {\bf 64}, 125010 (2001)
[arXiv:hep-ph/0003186].
;
R.~Arnowitt, B.~Dutta and Y.~Santoso,
Int.\ J.\ Mod.\ Phys.\ A {\bf 16S1B}, 852 (2001)
[arXiv:hep-ph/0011196].
;
J.~L.~Feng, K.~T.~Matchev and F.~Wilczek,
Phys.\ Lett.\ B {\bf 482}, 388 (2000)
[arXiv:hep-ph/0004043].
;
A.~Bottino, F.~Donato, N.~Fornengo and S.~Scopel,
Phys.\ Rev.\ D {\bf 63}, 125003 (2001)
[arXiv:hep-ph/0010203].
;
T.~Nihei, L.~Roszkowski and R.~Ruiz de Austri,
JHEP {\bf 0207}, 024 (2002)
[arXiv:hep-ph/0206266].


\bibitem{genius}
H.~V.~Klapdor-Kleingrothaus {\it et al.}  [GENIUS Collaboration],
arXiv:hep-ph/9910205

\bibitem{zeplin}
N.J.T. Smith et.al., 
Published in {\it Buxton 1998, The identification of dark matter} 335-340;
D.~Cline {\it et al.},
Astropart.\ Phys.\  {\bf 12}, 373 (2000).


\bibitem{dama}
R.~Bernabei {\it et al.}  [DAMA Collaboration],
.Phys. Lett.{\bf B480},23(2000)


\bibitem{cdms}
R. Abusaidi et.al., Phys. Rev. Lett.{\bf84}, 5699(2000),
"Exclusion Limits on WIMP-Nucleon Cross-Section
from the Cryogenic Dark Matter Search", CDMS Collaboration preprint
CWRU-P5-00/UCSB-HEP-00-01 and astro-ph/0002471.


\bibitem{edelweiss}
A.~Benoit {\it et al.},
arXiv:astro-ph/0206271.








\end{thebibliography}
\end{document}